\documentclass[letterpaper, 10 pt, conference]{ieeeconf}  

\IEEEoverridecommandlockouts                              

\usepackage{graphicx}

\title{\LARGE \bf
Robust Modelling of Reflectance Pulse Oximetry for SpO$_2$ Estimation
}

\author{Sricharan Vijayarangan$^{1*}$, Prithvi Suresh$^{1*}$, Preejith SP$^{1}$, Jayaraj Joseph$^{1}$ and Mohansankar Sivaprakasam$^{1,2}$  
\thanks{* Denotes Equal Contribution}
\thanks{$^{1}$ are with Healthcare Technology and Innovation Center (HTIC),
        Indian Institute of Technology (IIT-M), India
        {\tt\small sricharanv@htic.iitm.ac.in}}%
\thanks{$^{2}$ is with Department of Electrical Engineering,
        Indian Institute of Technology, Madras (IITM), India
        {}}%
}

\begin{document}

\maketitle
\thispagestyle{empty}
\pagestyle{empty}

\begin{abstract}
Continuous monitoring of blood oxygen saturation levels is vital for patients with pulmonary disorders. Traditionally, SpO$_2$ monitoring has been carried out using transmittance pulse oximeters due to its dependability. However, SpO$_2$ measurement from transmittance pulse oximeters is limited to peripheral regions. This becomes a disadvantage at very low temperatures as blood perfusion to the peripherals decreases. On the other hand, reflectance pulse oximeters can be used at various sites like finger, wrist, chest and forehead. Additionally, reflectance pulse oximeters can be scaled down to affordable patches that do not interfere with the user's diurnal activities. However, accurate SpO$_2$ estimation from reflectance pulse oximeters is challenging due to its patient dependent, subjective nature of measurement. Recently, a Machine Learning (ML) method was used to model reflectance waveforms onto SpO$_2$ obtained from transmittance waveforms. However, the generalizability of the model to new patients was not tested. In light of this, the current work implemented multiple ML based approaches which were subsequently found to be incapable of generalizing to new patients. Furthermore, a minimally calibrated data driven approach was utilized in order to obtain SpO$_2$ from reflectance PPG waveforms. The proposed solution produces an average mean absolute error of 1.81\% on unseen patients which is well within the clinically permissible error of 2\%. Two statistical tests were conducted to establish the effectiveness of the proposed method.  
\newline
\indent \textit{Clinical relevance}— The proposed method ameliorates our current understanding of reflectance based pulse oximetry and provides a method to estimate SpO$_2$ from reflectance pulse oximeters.
\end{abstract}

\section{INTRODUCTION}
Pulse oximetry is a non intrusive method used to measure the amount of oxygen saturation in a person’s blood. Peripheral Capillary Oxygen Saturation (SpO$_2$), an estimate of the amount of oxygen in the blood, is conventionally obtained by taking the ratio of volume of oxygenated haemoglobin to that of the total haemoglobin. This is estimated by passing light of two different wavelengths, generally Red (660 nm) and Infrared (IR)(890 nm), and analyzing the amount of light absorbed along the way  \cite{mortz1999system}. This provides information about the pulsatile arterial blood flow.

The most prevalent type of pulse oximetry is transmittance pulse oximetry, wherein the light passes through the region of interest (traditionally the finger) and the required analysis is carried out. A relatively unconventional method is reflectance pulse oximetry \cite{eidelman1999reflectance}, in which the intensity of the reflected light is analysed. Under a controlled clinical environment as described in P{\"a}lve \textit{et al.} \cite{palve1992reflection} and K{\"o}nig \textit{et al.} \cite{konig1998reflectance}, the need and success of reflectance pulse oximetry over transmittance is well established.

Despite its clinical necessity, estimation of SpO$_2$ from a reflectance photoplethysmogram requires a novel approach due to the inherent path difference between the red and IR wavelengths at the receiver. The variation in this path difference is not constant and depends upon the nature of one’s skin and tissue\cite{mendelson1988noninvasive}. This invalidates the use of one-time calibration as done in transmittance pulse oximeters. As a result, a novel data driven approach to model reflectance pulse oximetry ,subsequently estimating SpO$_2$, would enable a flexible yet robust approach that circumvents the current problems experienced by reflectance pulse oximetry

Venkat \textit{et al.} \cite{venkat2019machine} proposed a calibration-less Machine Learning (ML) approach to derive SpO$_2$ from reflectance PPG waveforms. However, the number of data points corresponding to each SpO$_2$ value obtained for the study is skewed in the favour of higher SpO$_2$ values, implying that the algorithm was not tested across all possible SpO$_2$ values. Furthermore, the proposed methodology was not tested on new patients. In order to arrive at a generalizable decision system centered around deriving SpO$_2$ from reflectance pulse oximetry, our contributions are as follows:

\begin{enumerate}
   \item We identified a data acquisition protocol that aims to acquire a wider range of values of SpO$_2$ from the same patient in order to understand the variation of reflectance PPG waveform with SpO$_2$.
   \item We perform a comprehensive analysis on the data acquired and corroborate the requirement of a data-driven approach.
   \item We examine previous ML methods and demonstrate its ineffectiveness in generalizing to unseen data. 
   \item We propose a method of estimation, requiring minimal calibration, that would generalize well to new patients. We quantify the estimation using Box plots and Bland Altman analysis.    
\end{enumerate}

\section{Design Methodology}
\subsection{Data Acquisition Protocol}
A clinical study was conducted to analyze the feasibility of estimating reflectance SpO$_2$ measurements from transmittance SpO$_2$. The hardware design used for all the experiments in the proposed method was the same as that used in Venkat \textit{et al.} \cite{venkat2019machine}. The device information, clinical protocols, and the study objectives were approved by the Institutional Review Board, Christian Medical College and Hospital (CMC), Vellore. Data was collected with informed consent from 28 subjects, suffering from various pulmonary disorders, with natural SpO$_2$ levels without external aid varying from 60\% to 100\%.  Reflectance probes were attached to the subject in the following locations: finger, wrist and forehead. Nellcor transmitance probe, which was used to obtain reference SpO$_2$, was attached to the finger (same hand) when the data was collected. 
\begin{figure}[t]
    \centering
    \includegraphics[width = 0.7\columnwidth]{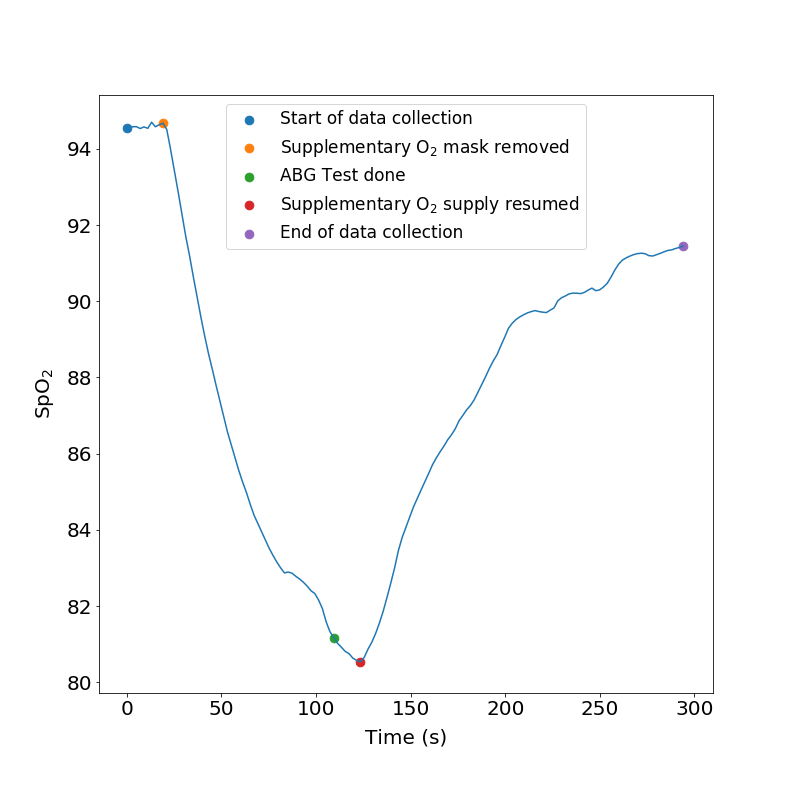}
    \setlength{\abovecaptionskip}{-10pt}
    \caption{Data Collection Procedure}
    \label{data_collect}
\end{figure}
Arterial Blood Gas (ABG) test \cite{davis2013aarc} was done on the subject, in parallel, to measure their SaO$_2$. Prior to the ABG test, oxygen mask was removed from the subject to measure their natural SpO$_2$ level. This period would correspond to a drop in the SpO$_2$ levels of the patient. Once the ABG test was conducted, the supplemental oxygen supply was resumed which led to an increase in SpO$_2$ levels. Note that the data collection (shown in Figure \ref{data_collect}) was initiated a few minutes before the ABG test was conducted and was concluded a few minutes after the test. To the best of our knowledge, this is the first procedure in which a continuous wide range of SpO$_2$ values were obtained for a single patient.   

\subsection{Data Preprocessing}
The transmittance and reflectance SpO$_2$ obtained from the fingers with a sampling frequency of 600 Hz were preprocessed in a similar fashion. These steps are elucidated below:
\begin{enumerate}
    \item \textbf{Moving Average (MAV) Filter}: The Red and IR signals were passed through a MAV filter with a window length of 50 samples to smoothen them, thus removing high frequency noise components.
    \item \textbf{Detrending}: Since both the red and IR signals had exhibited baseline wander, detrending was carried out to ease the process of peak and valley detection. 
    \item \textbf{Peak and valley Detection}: Position of peaks and valleys were extracted from the detrended signals. Subsequently, erroneous peaks and valleys were discarded based on the average peak-to-peak and valley-to-valley distance.
    \item \textbf{Signal Quality Check}: The signals were deemed fit based on the cross correlation between the red and IR signals. 
\end{enumerate}
\renewcommand{\thefootnote}{\fnsymbol{footnote}}
The length of the window and the percentage overlap was empirically chosen to be 4 seconds and 25\% respectively. This was decided based on the average change in the DC value within a window.\footnote[2]{ code is available at https://github.com/prithusuresh/Reflectance-SPO2}
\subsection{Problem Formulation}
From each window in the transmittance waveform, a ratio was obtained by combining AC and DC components of both wavelengths, using the following formula 
\begin{center}
    $R\_value = \frac{(RED_{AC}/RED_{DC})}{(IR_{AC}/IR_{DC})}$ 
\end{center}
The actual SpO$_2$ was obtained from the $R\_value$, which is the standard method for SpO$_2$ estimation. The goal of the proposed method is to estimate the SpO$_2$ value by modelling the relationship between the reflectance PPG waveform and the transmittance SpO$_2$.
\begin{figure}[t]
    \centering
    \includegraphics[width = 0.50\columnwidth]{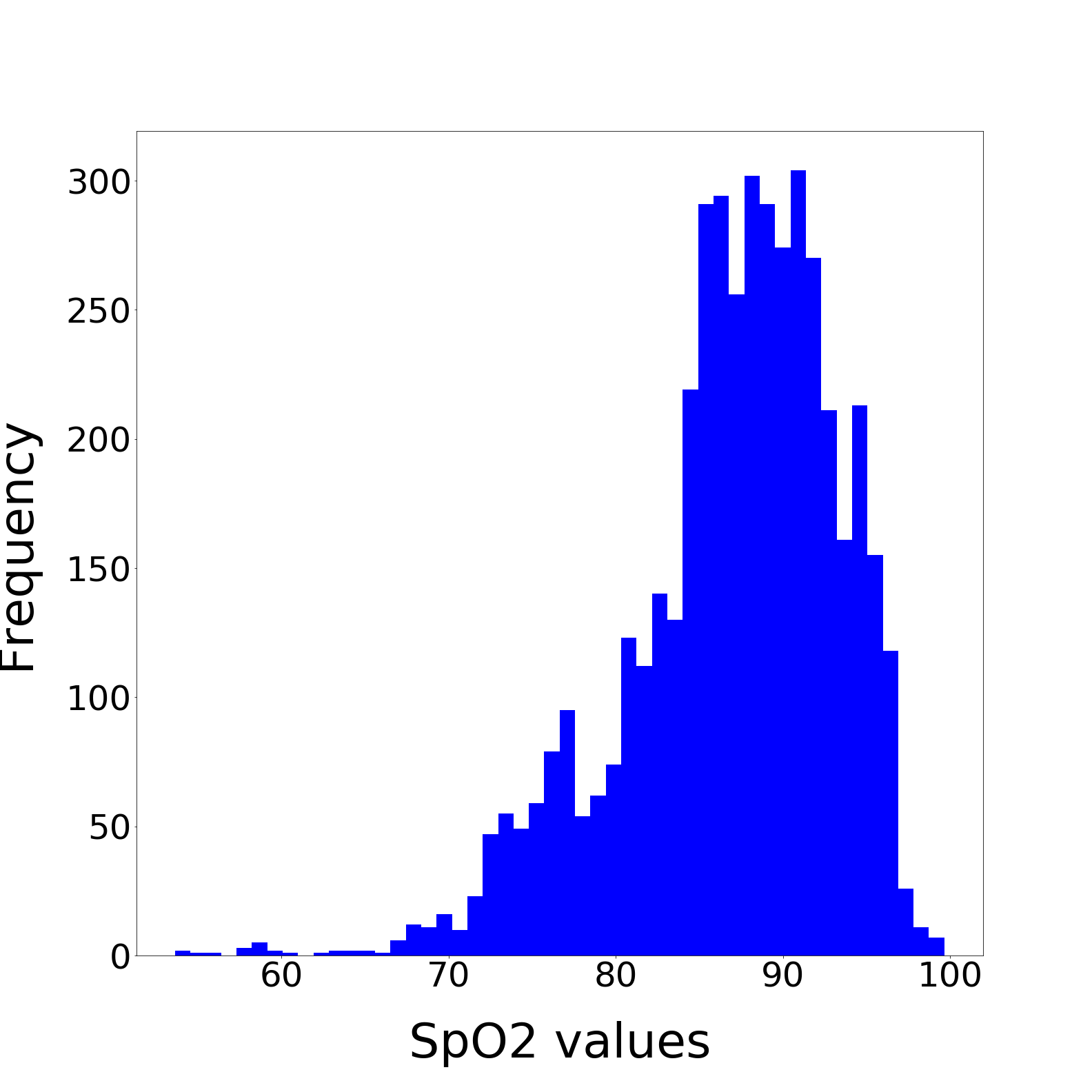}
    \caption{Distribution of SpO$_2$ for overall data collected}
    \label{hist}
\end{figure}

\subsection{Exploratory Data Analysis (EDA)}
\label{eda}
Although the data acquisition obtained a wide range of SpO$_2$ values across various patients, it is clear from Figure \ref{hist} that this approach does not reduce the massive imbalance against the lower values($<$75\%) of SpO$_2$. This was expected as it is prohibitively hard to allow a patient to have a low SpO$_2$ level for longer periods.

\begin{figure}[t]
    \centering
    \includegraphics[width= 0.55\columnwidth]{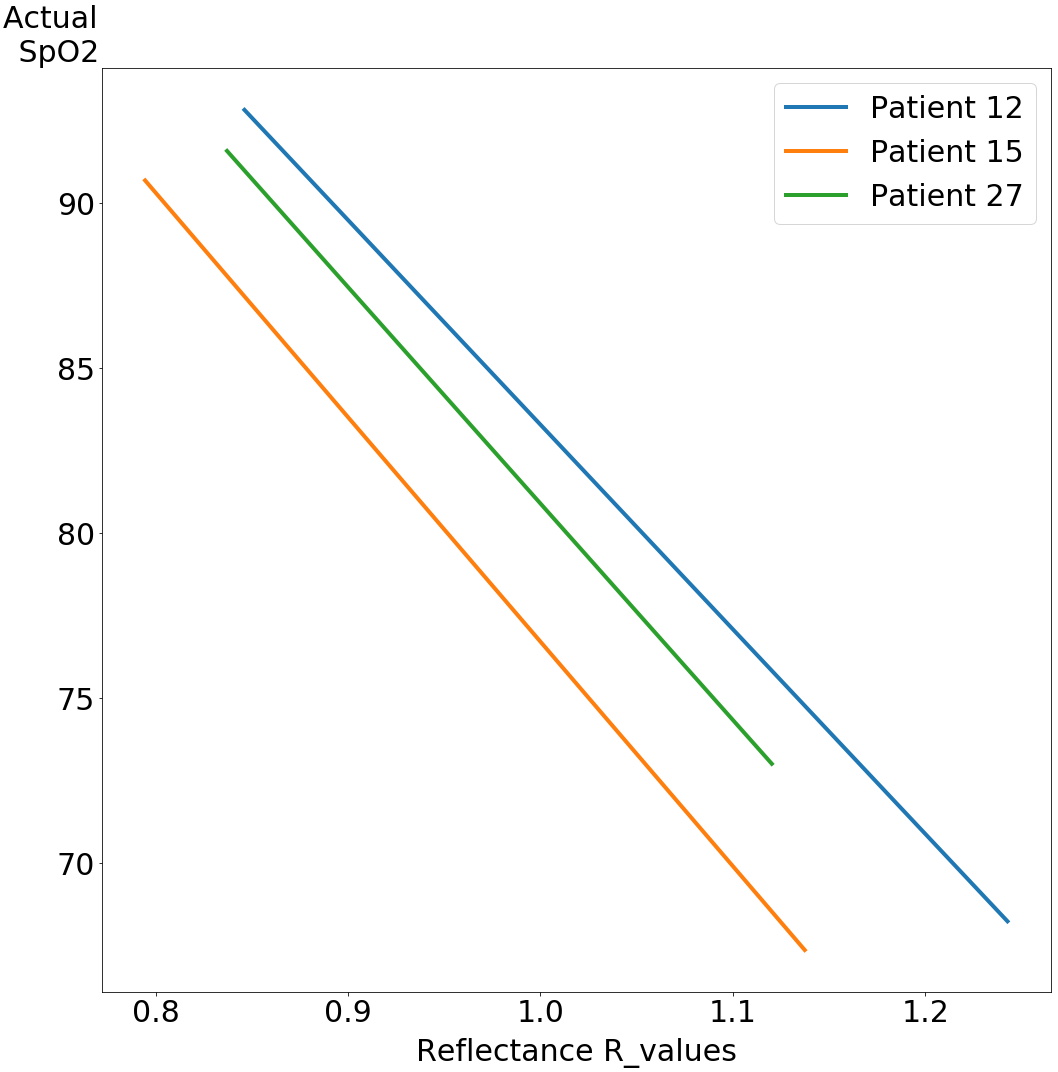}
    \caption{Trends for SpO$_2$ vs R for three patients selected at random from the train set. Note that these trends were obtained by fitting a line to the scatter plot}.
    \label{Trendlines}
\end{figure}

Furthermore, a linear relationship between $R\_value$, calculated from the reflectance signal (reflectance-$R\_value$), and the corresponding SpO$_2$, obtained from the transmittance signal (transmittance-SpO$_2$), was observed for individual patients as shown in Figure \ref{Trendlines}. For each individual a unique straight line was obtained. Examples of this trend is shown on three patients who were randomly picked. However, there were two patients whose $R\_value$-SpO$_2$ relationship deviated from the observation above. On closer examination of their PPG waveform, it was inferred that different reflectance-$R\_value$s were getting mapped to the same transmittance-SpO$_2$ values. The reason for the same is discussed in Section \ref{discussion}.

\section{Experiments and Results}

\subsection{Examination of previous machine learning methods}
Venkat \textit{et al.} \cite{venkat2019machine} proposed to model reflectance pulse oximetry by using a classifier wherein each target class corresponds to a rounded off SpO$_2$ value. The data representing each SpO$_2$ class was split into a train-validation set of 70\%-30\%, which was later accumulated. This was done to tackle the severe imbalance in the representation of lower SpO$_2$ values. Venkat \textit{et al.} showed that the bagging of decision trees provides the best accuracy across different values of SpO$_2$. In order to test the ability of these models to generalize, the current work splits the data into three sets namely, train, validation and test sets. To deal with the imbalance, majority undersampling was utilized. Data obtained from 20 patients was split randomly into 10 folds of train and validation. Data from the rest of the patients was put into a test set. Both bagging (Random Forest Classifier and Random Forest Regressor) and boosting (Gradient boosting) were tried. Multiple features as given in Venkat \textit{et al.} \cite{venkat2019machine} were extracted from the transmittance waveform. The reported evaluation metrics are Average Mean Square Error (Avg MSE), Average Mean Absolute Error (Avg MAE) and the Average R$^2$ Score (Avg R$^2$ Score) across the validation and test set. Table \ref{ml_method} highlights the performance of multiple machine learning models on the validation set. Table \ref{ml_method_2} highlights the performance of all these models on the test set. 

\subsection{Proposed Estimation Method}
As the transmittance-SpO$_2$ vary linearly with the reflectance-$R\_values$  (corroborated in Figure \ref{Trendlines}), we use a straight line to model the relationship between transmittance-SpO$_2$ and reflectance-$R\_value$. Given a new patient, whose reflectance-$R\_value$ and transmittance-SpO$_2$ are known, the objective is now reduced to finding the best fit line. 
The training set was computed from the data of those patients ($\Pi$), whose SpO$_2$ values varied over a range greater than 15, which totalled to 10 patients. The training set ($\tau$) consists of individual straight lines ($l_i$) that best captured their respective  reflectance-$R\_value$ and transmittance-SpO$_2$ relationship. Thus,
\begin{center}
    $\tau$ = \{$l_i$\} $\forall\ i\ \in\ \Pi $  
\end{center}

The calibration set ($\gamma$) consists of five values of reflectance-$R\_value$ and transmittance-SpO$_2$ pairs. The number of points chosen in the calibration set is a trade off between an accurate representation of the new patient versus the calibration time. All these five points were values obtained between 90-95, as these are better represented in our dataset Figure \ref{hist}. The lateral distance ($LD$), defined as the absolute difference in $R\_value$ between each point in the $\gamma$  and each line in $\tau$ for the corresponding SpO$_2$, is computed. A point is matched to a line in $\tau$ if it has the least $LD$. The line $l_i$, with the most number of matches, is selected to model the new patient. 
Table \ref{tab:Calibration_Results} highlights the results obtained using the proposed decision system on twelve unseen patients. Only these patients exhibited a SpO$_2$ range over 10\%. The other 6 patients displayed an inadequate range and were not considered.   

\begin{table}[t]
\caption{Performance of ML models on Validation set}
\begin{center}
\begin{tabular}{|l|l|l|l|}
\hline
Model                  & Avg MSE     & Avg MAE     &  Avg R$^2$ Score \\ \hline
RF Regressor      & 1.284       & 1.284       & 0.941            \\ \hline
RF Classifier     & 0.941       & 0.580       & 0.928            \\ \hline
Gradient Boosting & 2.617       & 1.150       & 0.762            \\ \hline
\end{tabular}
\end{center}
\label{ml_method}
\end{table}


\begin{table}[t]
\caption{Performance of ML models on test set}
\begin{center}
\begin{tabular}{|l|l|l|l|}
\hline
Model                  & Avg MSE & Avg MAE & Avg R$^2$ Score \\ \hline
RF Regressor      & 8.213   & 6.557   & -2.901       \\ \hline
RF Classifier     & 8.213   & 6.557   & -2.901       \\ \hline
Gradient Boosting & 5.175   & 4.490   & -0.550       \\ \hline
\end{tabular}
\end{center}
\label{ml_method_2}
\end{table}

\begin{table}[b]
\center
\caption{Mean Absolute errors across all patients}
\begin{tabular}{|l|l|}
\hline
Patient Id & MAE \\ \hline & \\[-1em]
1          & 3.18    \\ \hline 
2          & 1.80    \\ \hline
3          & 3.28    \\ \hline 
4          & 1.40    \\ \hline 
5          & 0.92    \\ \hline 
6          & 2.79    \\ \hline 
7          & 1.08    \\ \hline 
8          & 1.55    \\ \hline
9          & 0.66    \\ \hline 
10         & 1.87    \\ \hline
11         & 1.75    \\ \hline 
12         & 1.39    \\ \hline
\end{tabular}
\label{tab:Calibration_Results}
\end{table}

\section{Discussion}
\label{discussion}

From Table \ref{ml_method}, it is observed that the ML methods do well on the validation set. This implies that the model seems to be performing well on the unseen SpO$_2$ values of the same patient. However, Table \ref{ml_method_2} shows that the ML models do not generalize well to the test set. The poor performance of these models can be explained by the inherent difference in paths of the both the red and IR wavelengths at the receiver's end for different patients. The path difference occurs as a result of difference in finger thickness, skin colour and other possible variances. These results show that the training set does not effectively capture all possible variations in SpO$_2$ and that the features that were used were not discriminative of the path length in different people.

\begin{figure}[t]
    \centering
    \includegraphics[width = 0.60\columnwidth]{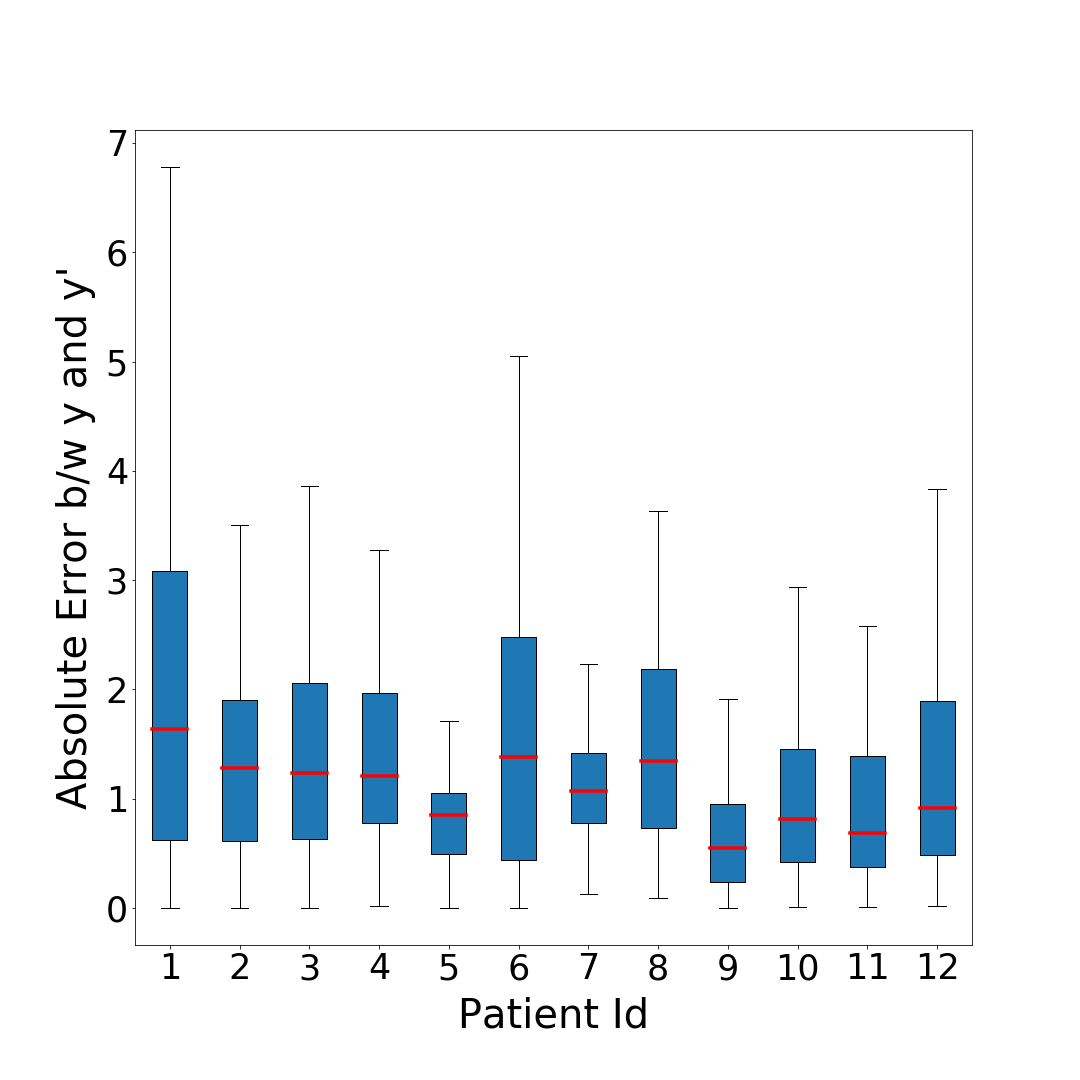}
    \caption{Box plot of MAE's for all windows across all patients in the unseen test set. y refers to actual SpO$_2$ while y' refers to the Predicted SpO$_2$}
    \label{fig:box}
\end{figure}

Table \ref{tab:Calibration_Results} demonstrates the effectiveness of the proposed method on the test set. The Mean Absolute Error (MAE) is computed for all the patients on the test set. The Average of the MAE across all patients is 1.81\%. This significantly outperforms the best ML model which gave an Average MAE of 4.490. However, for certain patients it was observed that the MAE values were as high as 3.18. On closer analysis, it was observed that this was due to predictions on certain windows being highly erroneous, resulting in a skewed average value. This aberration is a result of abnormal $R\_value$ computation due to a high DC component of the red wavelength. The high DC component caused mutliple reflectance-$R\_value$s to be mapped to the same SpO$_2$ values, as observed in section \ref{eda}. This observation can be corroborated by the box plot shown in Figure \ref{fig:box}. It can be seen that the 75\% quartile range lies within 2\% for all except one patient. The few outliers had produced a large error thus increasing the average of the MAE. However, the box plot does not take into account the possibility of systematic and random errors or bias in the measurement and computation of SpO$_2$ from the transmittance PPG waveform.
 
\begin{figure}[t]
    \centering
    \includegraphics[width=1\columnwidth]{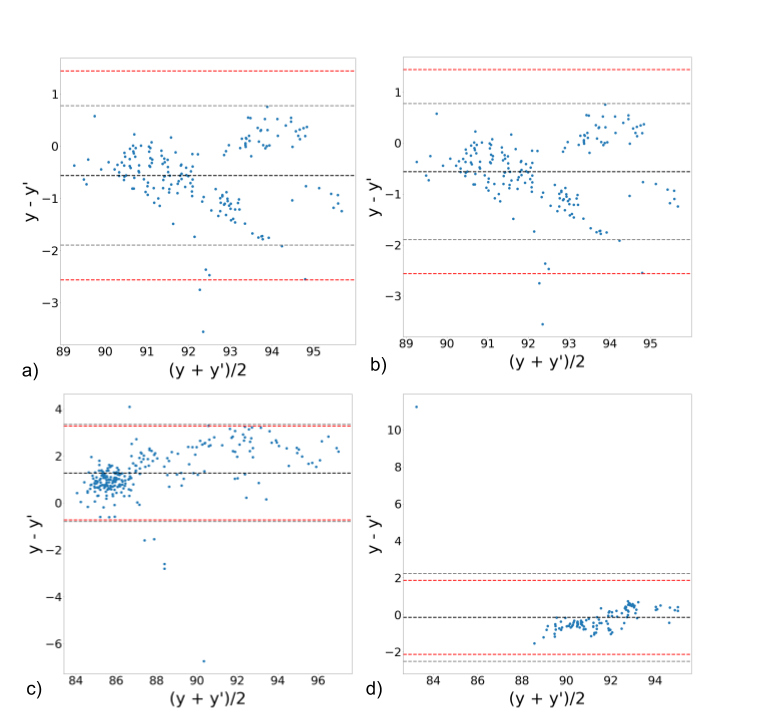}
    \setlength{\abovecaptionskip}{-20pt}
    \caption{Bland Altman computed across all patients. y refers to the Actual SpO$_2$ value and y' refers to the predicted SpO$_2$ value. The black,red and grey lines correspond to the means of differences, clinical threshold and the limits of agreement, respectively.}
    \label{fig:Bland_Altman}
\end{figure}

A better understanding about the degree of agreement between the estimated SpO$_2$ and the actual SpO$_2$ that accounts for the errors in measurement of SpO$_2$ is given by the Bland-Altman analysis. To this end, the Bland-Altman plots for four patients are shown in Figure \ref{fig:Bland_Altman}. The Limits of Agreement (LoA) were defined as $\pm1.96SD$ of the mean difference. Generally, it is recommended that the accurately predicted observations lie within the error band of $\pm2\%$.  Figure \ref{fig:Bland_Altman}a and \ref{fig:Bland_Altman}b show examples where the LoA is well within the error band indicating good estimations of SpO$_2$. It is also clear that more than 95\% of the points are within the LoA. These plots are representative of 83\% of the patients. For two patients, due to the presence of large outliers, the LoA marginally eclipses the error band as shown in Figure \ref{fig:Bland_Altman}c and \ref{fig:Bland_Altman}d. Given more training data that covers a larger distribution and a robust preprocessing stage to discard signals with abnormally high values of DC amplitudes, the proposed method is likely to generalize even better to unseen data. Nevertheless, it provides a massive improvement over ML based methods across all patients. 

\section{CONCLUSIONS}
This work proposes a novel method for estimating SpO$_2$ from reflectance pulse oximetry with minimal calibration using SpO$_2$ obtained from transmittance PPG signals. However, subsequent statistical analysis showed that the modelling could be improved with more data and better selectivity. 

\section{ACKNOWLEDGEMENT}
The authors would like to acknowledge Dr Balamugesh and Dr Christopher DJ, Department of pulmonary medicine, Christian Medical College and Hospital (CMC), Vellore, who provided valuable expertise regrading the data and its acquisition process. 

\bibliographystyle{IEEEtran}
\bibliography{root}

\begin{thebibliography}{1}
\providecommand{\url}[1]{#1}
\csname url@samestyle\endcsname
\providecommand{\newblock}{\relax}
\providecommand{\bibinfo}[2]{#2}
\providecommand{\BIBentrySTDinterwordspacing}{\spaceskip=0pt\relax}
\providecommand{\BIBentryALTinterwordstretchfactor}{4}
\providecommand{\BIBentryALTinterwordspacing}{\spaceskip=\fontdimen2\font plus
\BIBentryALTinterwordstretchfactor\fontdimen3\font minus
  \fontdimen4\font\relax}
\providecommand{\BIBforeignlanguage}[2]{{%
\expandafter\ifx\csname l@#1\endcsname\relax
\typeout{** WARNING: IEEEtran.bst: No hyphenation pattern has been}%
\typeout{** loaded for the language `#1'. Using the pattern for}%
\typeout{** the default language instead.}%
\else
\language=\csname l@#1\endcsname
\fi
#2}}
\providecommand{\BIBdecl}{\relax}
\BIBdecl

\bibitem{mortz1999system}
M.~S. Mortz, ``System for pulse oximetry spo2 determination,'' Aug.~10 1999, uS
  Patent 5,934,277.

\bibitem{eidelman1999reflectance}
A.~I. Eidelman, Y.~Saroussi, and I.~Y. Shemesh, ``Reflectance pulse oximetry
  (rpo): A new tool for continous measurement of heart rate and oxygen
  saturation,'' \emph{Pediatric Research}, vol.~45, no.~2, pp. 16--16, 1999.

\bibitem{palve1992reflection}
H.~P{\"a}lve, ``Reflection and transmission pulse oximetry during compromised
  peripheral perfusion,'' \emph{Journal of clinical monitoring}, vol.~8, no.~1,
  pp. 12--15, 1992.

\bibitem{konig1998reflectance}
V.~K{\"o}nig, R.~Huch, and A.~Huch, ``Reflectance pulse oximetry--principles
  and obstetric application in the zurich system,'' \emph{Journal of Clinical
  Monitoring and Computing}, vol.~14, no.~6, pp. 403--412, 1998.

\bibitem{mendelson1988noninvasive}
Y.~Mendelson and B.~D. Ochs, ``Noninvasive pulse oximetry utilizing skin
  reflectance photoplethysmography,'' \emph{IEEE Transactions on Biomedical
  Engineering}, vol.~35, no.~10, pp. 798--805, 1988.

\bibitem{venkat2019machine}
S.~Venkat, M.~T.~A. PS, A.~Alex, S.~Preejith, D.~Christopher, J.~Joseph,
  M.~Sivaprakasam \emph{et~al.}, ``Machine learning based spo2 computation
  using reflectance pulse oximetry,'' in \emph{2019 41st Annual International
  Conference of the IEEE Engineering in Medicine and Biology Society
  (EMBC)}.\hskip 1em plus 0.5em minus 0.4em\relax IEEE, 2019, pp. 482--485.

\bibitem{davis2013aarc}
M.~D. Davis, B.~K. Walsh, S.~E. Sittig, and R.~D. Restrepo, ``Aarc clinical
  practice guideline: blood gas analysis and hemoximetry: 2013,''
  \emph{Respiratory care}, vol.~58, no.~10, pp. 1694--1703, 2013.

\end{thebibliography}

\end{document}